\documentclass[a4paper,11pt]{article}
\usepackage{pos}

\def\powhegbox{\textsc{Powheg-Box}}
\def\mg5{\textsc{MG5\_aMC@NLO}}

\title{Comparison of $t\bar{t}W$ theory predictions in the $3\ell$ channel}

\author[a]{Manfred Kraus}

\affiliation[a]{Physics Department, Florida State University,\\
Tallahassee, FL 32306-4350, USA}

\emailAdd{mkraus@hep.fsu.edu}

\abstract{
We report on our recent comparison of various theoretical approaches to predict
fiducial signatures for $pp\to t\bar{t}W$ in the $3\ell$ decay channel at
$\mathcal{O}(\alpha_s^3\alpha^6)$ and $\mathcal{O}(\alpha_s\alpha^8)$. The
comparison includes fixed-order predictions including full off-shell effects as
well as predictions based only on the double-resonant contributions by
employing the Narrow-Width-Approximation. Furthermore, we include parton-shower
matched predictions using the \mg5{} and \powhegbox{} frameworks.
}

\FullConference{%
  41st International Conference on High Energy physics - ICHEP2022\\
  6-13 July, 2022\\
  Bologna, Italy
}

\begin{document}
\maketitle

\section{Introduction}
The production of top-quark pairs in association with a $W$ boson is one of the
rarest processes in the Standard Model. At the same time, it gives rise to a
multitude of decay signatures of unfathomable complexity. The $pp\to t\bar{t}W$
process is one of the main backgrounds in $t\bar{t}H$ measurements and searches
for the $t\bar{t}t\bar{t}$ process. Therefore, a precise understanding of the
$pp\to t\bar{t}W$ process is inevitable. This is even more emphasized as recent
measurements of the $t\bar{t}W$ component as part of the $t\bar{t}H$ analysis
show tensions~\cite{ATLAS:2019nvo,CMS:2020iwy} with the Standard Model
predictions.

Due to its importance, the $pp\to t\bar{t}W$ process has received plenty of
attention in the theory community. First predictions at NLO QCD accuracy for
production and decay have been reported in Ref.~\cite{Campbell:2012dh}.
Subsequently, NLO EW corrections for on-shell $t\bar{t}W$ production have been
computed for the first time in Refs.~\cite{Frixione:2015zaa,Frederix:2018nkq}.
Furthermore, mixed QCD and EW contributions have been studied in
Refs.~\cite{Dror:2015nkp, Frederix:2017wme}. Also the effects of soft-gluon
resummation have been studied in detail in Refs.~\cite{Li:2014ula,
Broggio:2016zgg,Kulesza:2018tqz,Broggio:2019ewu}.  In order to describe
fiducial signatures the on-shell $pp\to t\bar{t}W$ process has been matched to
parton showers using either the MC@NLO~\cite{Maltoni:2014zpa,Maltoni:2015ena,
Frederix:2020jzp} or the POWHEG~\cite{Garzelli:2012bn, FebresCordero:2021kcc}
approach. Further higher-order corrections have also been included via
multi-jet merging~\cite{vonBuddenbrock:2020ter,Frederix:2021agh}.  An
orthogonal approach to describe fiducial signatures are fixed-order
computations based on matrix elements for the fully decayed process, e.g. $pp
\to e^+\nu_e\mu^-\bar{\nu}_\mu e^+\nu_e b\bar{b}+X$, which ultimately account
for all double, single and non-resonant top-quark contributions. For the three
lepton decay channel both, NLO QCD predictions~\cite{Bevilacqua:2020pzy,
Denner:2020hgg,Bevilacqua:2020srb} as well as EW contributions
\cite{Denner:2021hqi,Bevilacqua:2021tzp} have been studied in the literature.

In this proceedings, we report on our latest study~\cite{Bevilacqua:2021tzp}
that aims at comparing parton-shower and fixed-order full off-shell
computations at the fiducial level. A detailed comparison of both approaches is
mandatory as they include very different aspects of physics but aim to describe
fiducial signatures accurately.

\section{Computational setup}
In our comparative study we employ the following computational approaches for the
$pp\to t\bar{t}W$ process in the three-lepton decay channel.
\begin{description}
 \item[full off-shell:] The calculation of the full off-shell process is
performed using \textsc{Helac-Nlo}~framework\cite{Bevilacqua:2011xh,
Ossola:2007ax,vanHameren:2009dr,Czakon:2009ss,Bevilacqua:2013iha,Czakon:2015cla}.
We use the full off-shell calculation of Refs.
\cite{Bevilacqua:2020pzy,Bevilacqua:2021tzp} for the $pp \to
\ell^+\nu\ell^-\bar{\nu} \ell^\pm\nu b\bar{b}$ process. In this calculation
unstable particles are described in the complex mass scheme by Breit-Wigner
propagators.Furthermore, the computation not only includes the double resonant
but also single and non-resonant contributions. Naturally in this approach, NLO
QCD corrections to top-quark decays are automatically included.

\item[NWA:] Based on our recent automation of the Narrow-width-approximation
(NWA) in our framework~\cite{Bevilacqua:2019quz} we provide also predictions
for on-shell $pp\to t\bar{t}W$ predictions including NLO QCD top-quark decays.
This is achieved by taking the limit
\begin{equation}
 \lim_{\Gamma/m} \frac{1}{(p^2-m^2)^2 + m^2\Gamma^2} = 
 \frac{\pi}{m\Gamma}\delta(p^2-m^2)\;,
\end{equation}
which allows to factorize the process into production and decays stages.
These predictions are employed in order to disentangle various effects such as
off-shell effects and the perturbative treatment of the top-quark decay.

 \item[\powhegbox:] We also employ the recent \powhegbox{}
implementation~\cite{FebresCordero:2021kcc,Honeywell:2018fcl,Figueroa:2021txg}
to generate parton-shower matched predictions for $pp\to t\bar{t}W$. Decays
are included at LO accuracy keeping approximately spin correlations.

 \item[\mg5:] Finally, we also employ \mg5{}~\cite{Alwall:2014hca} in
conjunction with \textsc{MadSpin}~\cite{Artoisenet:2012st} to obtain a second
set of parton shower matched predictions. These have the same formal accuracy
as the \powhegbox{} results.
\end{description}
In the case of \powhegbox{} and \mg5{} predictions we employ the
\textsc{Pythia8}~\cite{Sjostrand:2014zea,Bierlich:2022pfr} parton shower, where
we neglect effects from hadronization and multiple parton scattering. For all
computational approaches we investigate theoretical uncertainties by means of
scale variations and if appropriate variation of matching scheme parameters.
For a more detailed account of the differences between the various approaches
as well as the computational setup refer the reader to
Ref.~\cite{Bevilacqua:2021tzp}.
\section{Phenomenological results}
We start the discussion of our findings at the level of inclusive cross
sections, since we can establish some global differences between the
computations already here. 
\begin{table}[h]
\centering
\begin{tabular}{c|cc}
 $\sigma^{\textrm{NLO}}_{\textrm{QCD}}$ & $t\bar{t}W$ QCD [fb] & $t\bar{t}W$ EW [fb] \\
\hline
full off-shell     & $1.58^{+3\%}_{-6\%}$   & $0.206^{+22\%}_{-17\%}$ \\
full NWA           & $1.57^{+3\%}_{-6\%}$   & $0.190^{+22\%}_{-16\%}$ \\
NWA with LO decays & $1.66^{+10\%}_{-10\%}$ & $0.162^{+22\%}_{-16\%}$ \\
\hline
\powhegbox{}       & $1.40^{+11\%}_{-11\%}$ & $0.133^{+21\%}_{-16\%}$ \\
\mg5{}             & $1.40^{+11\%}_{-11\%}$ & $0.136^{+21\%}_{-16\%}$
\end{tabular}
\caption{Inclusive cross sections for $t\bar{t}W$ QCD
$(\mathcal{O}(\alpha_s^3\alpha^6)$ and for $t\bar{t}W$ EW
$(\mathcal{O}(\alpha_s\alpha^8)$ at NLO QCD accuracy for various computational
approaches.}
\label{tab:xsec}
\end{table}
In Tab.~\ref{tab:xsec} the inclusive fiducial cross sections including the
estimated theoretical uncertainties are shown for the five different
calculations employed in our study. First, we observe that the subleading EW
contributions amount to roughly $13\%$ of the leading QCD cross section. 
Furthermore, we notice that in the case of the EW production mode the difference
between the full off-shell and the full NWA calculation is of the order of
$9\%$.  This is surprisingly large because these effects are expected to be of
the order of $\Gamma_t/m_t \sim 0.8\%$. We also find that the theoretical
uncertainty of the dominant QCD contribution is reduced if NLO QCD corrections
for the top-quark decays are included. Contrary, we do not observe this trend
in the EW case, as the corrections are dominated by the $pp\to t\bar{t}Wj$
production matrix elements. Finally, we find that the NLO+PS predictions are in
very good agreement with each other but yield a $11-34\%$ reduced cross section
with respect to the full off-shell calculation. The origin of this reduction is
due to multiple radiation in the resonant top-quark decays during the shower
evolution.

Let us now turn to the discussion of differential cross sections.
\begin{figure}
 \includegraphics[width=0.48\textwidth]{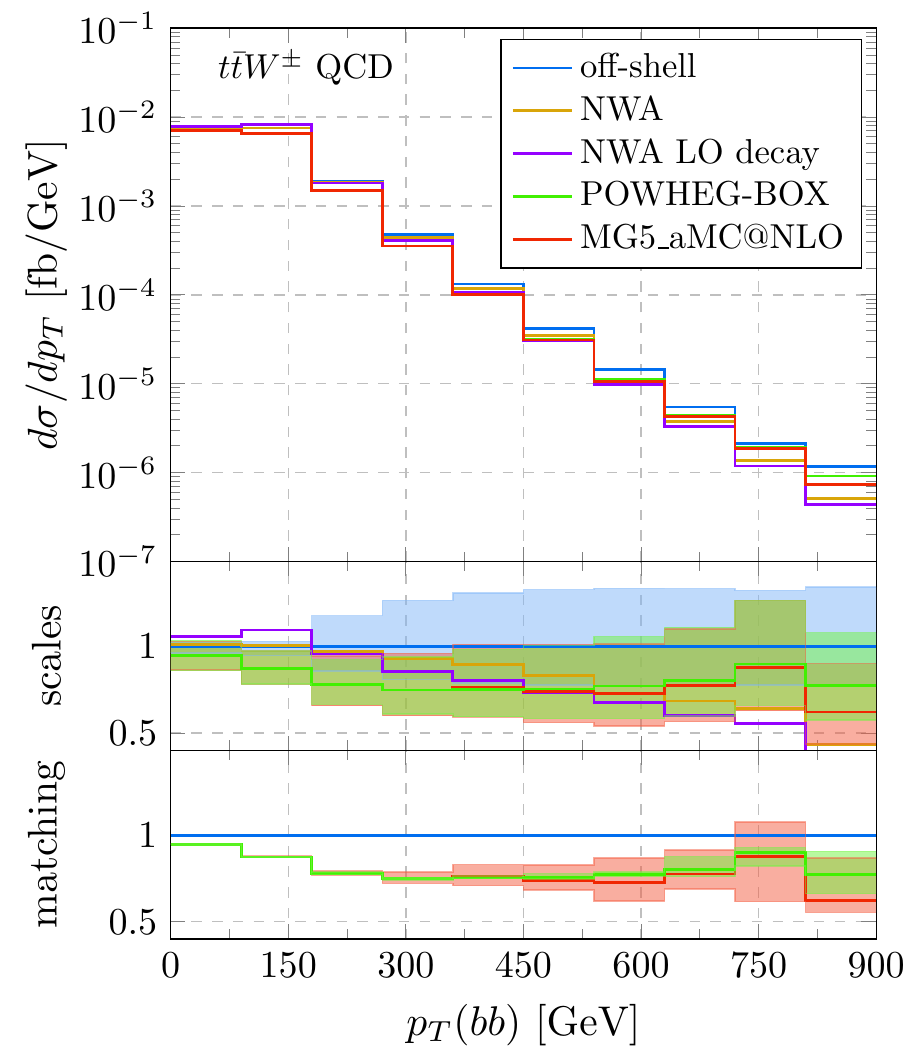}
 \includegraphics[width=0.48\textwidth]{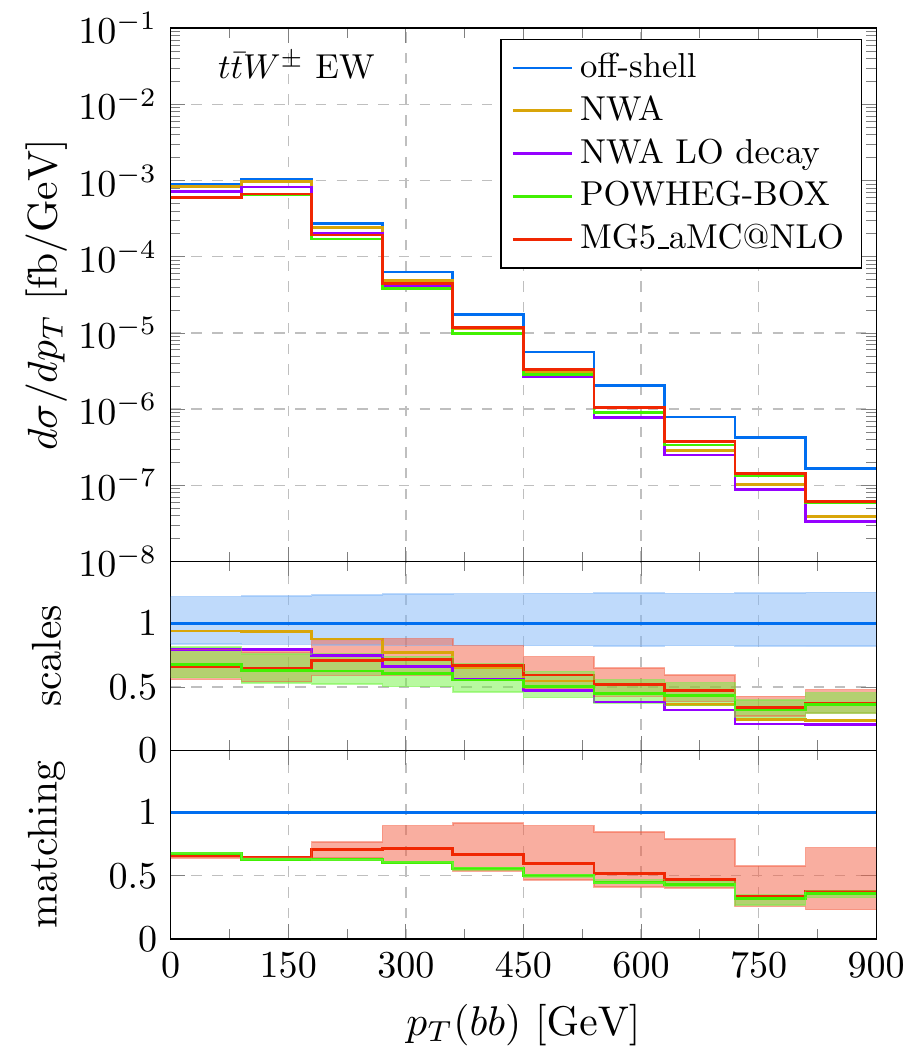}
 \caption{Differential distribution of the transverse momentum of the two
hardest $b$ jets for $t\bar{t}W$ QCD (left) and $t\bar{t}W$ EW (right). Figure
taken from Ref.~\cite{Bevilacqua:2021tzp}.} 
\label{fig:res_1}
\end{figure}
As an example, we present in Fig.~\ref{fig:res_1} the transverse momentum
distribution of the two hardest $b$ jets. On the left, the dominant $t\bar{t}W$
QCD predictions are shown, while on the right the subleading $t\bar{t}W$ EW
ones. The upper panels show the absolute predictions, the middle ones shows the
theoretical uncertainties estimated by independent scale variations normalized
to the full off-shell prediction and the bottom panel depicts matching
uncertainties also normalized to the off-shell calculation.

For the $t\bar{t}W$ QCD predictions on the right, we observe that the NWA is a
very good approximation of the full off-shell calculation in the bulk of the
distribution. Only in the tail of the spectrum considerable deviations are
visible. The parton shower predictions, on the other hand, have a very
different shape over the whole range of the distribution.  Nonetheless, all
generators are consistent with each other within the estimated uncertainties.
The theoretical uncertainties are also dominated by missing higher-order
corrections.
In contrast, the $t\bar{t}W$ EW contributions, shown on the right plot, show a
very different behavior. Not even the NWA performs well in this case. For
transverse momenta larger than roughly $450$ GeV all predictions deviate more
than $50\%$ from the full off-shell calculation. The predictions also become
quickly incompatible with each other within the uncertainties. The exception is
$\mg5$ as its uncertainties are severely inflated due to matching
uncertainties.

\begin{figure}
 \includegraphics[width=0.48\textwidth]{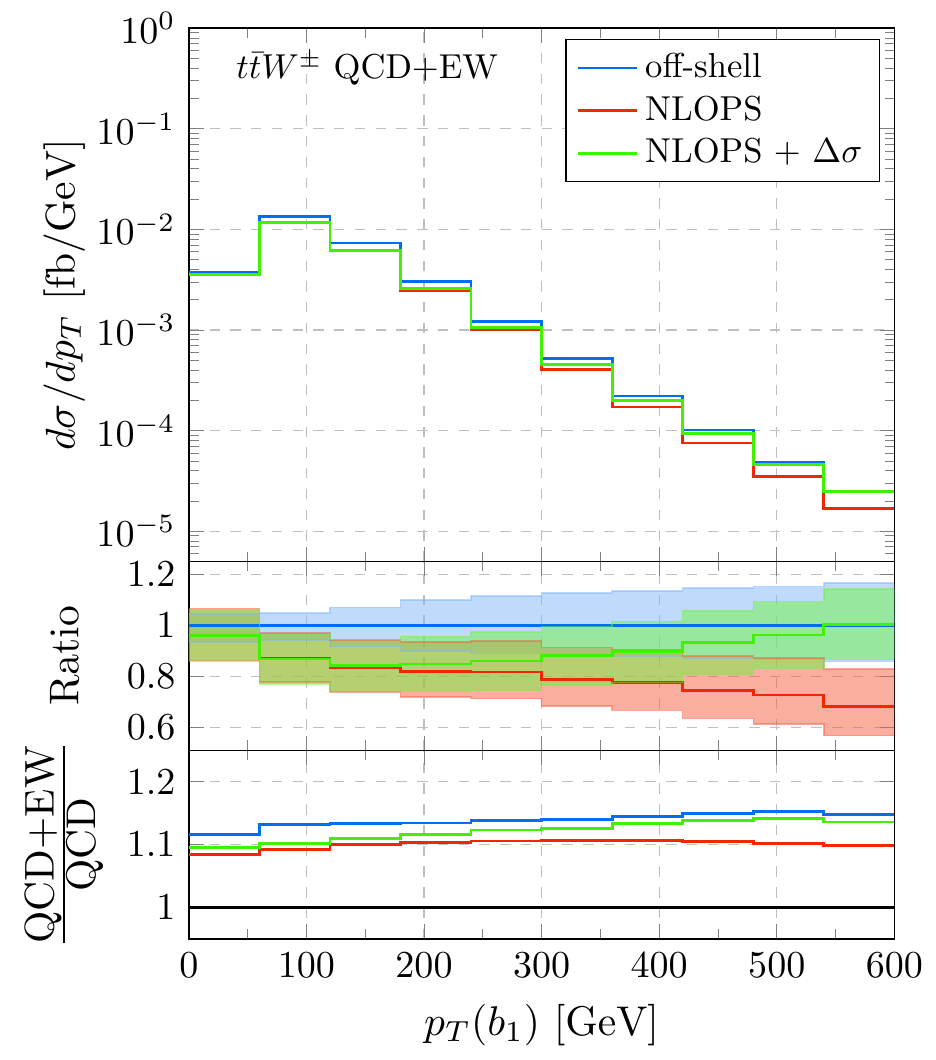}
 \includegraphics[width=0.48\textwidth]{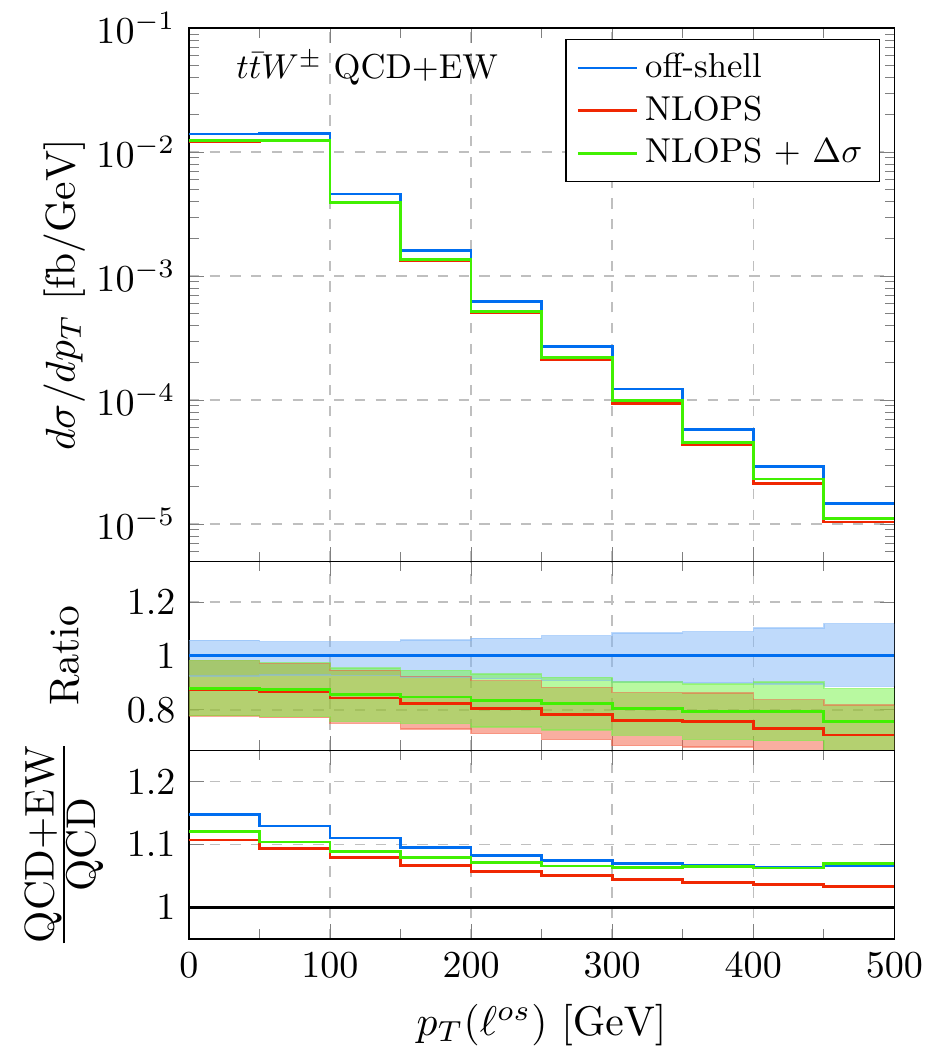}
 \caption{Differential distribution of the transverse momentum of the hardest
$b$ jet (left) and of the opposite-sign lepton $\ell^{os}$ (right). Figure
taken from Ref.~\cite{Bevilacqua:2021tzp}.}
 \label{fig:res_2}
\end{figure}
As the full off-shell calculation is not yet matched with parton showers we
propose to improve the currently available on-shell NLOPS calculations by a
simple procedure. We add off-shell corrections to NLOPS predictions via 
\begin{equation}
 \frac{d\sigma^{\textrm{th}}}{dX} = \frac{d\sigma^{\textrm{NLOPS}}}{dX} + 
 \frac{d\Delta\sigma_{\textrm{off-shell}}}{dX}\;, \qquad
 \frac{d\Delta\sigma_{\textrm{off-shell}}}{dX} = 
 \frac{d\sigma^{\textrm{NLO}}_{\textrm{off-shell}}}{dX} - 
 \frac{d\sigma^{\textrm{NLO}}_{\textrm{NWA}}}{dX}\;.
\end{equation}
The definition of $\Delta\sigma_{\textrm{off-shell}}$ removes approximately the
double counting between the double-resonant $t\bar{t}W$ contributions. It,
therefore, adds single and non-resonant contributions as well as interference
effects. The impact of these corrections are shown in Fig.~\ref{fig:res_2}, where
on the left the transverse momentum of the leading $b$ jet and on the right of
the opposite-sign lepton $\ell^{os}$ is shown. In the case of $p_T(b_1)$, we
find that the off-shell corrections are sizable in the tail of the
distribution. This is expected as this phase space region is dominated by
associated single-top production. In addition, we observe that the EW
contributions receive sizable corrections. However, the combined predictions,
NLOPS + $\Delta\sigma$, reproduce the tails of the full off-shell predictions to
a very good extent.

On the other hand, for $p_T(\ell^{os})$ we find only minor corrections. The
reason for this is that the distribution is described in an excellent way by
the NWA. Therefore, we obtain only very small corrections $d\Delta\sigma/dX$
over the whole plotted range. The residual corrections originate from the EW
contributions as can be deduced from the bottom panel. The two shown
differential distributions illustrate that the $\Delta\sigma$ correction terms
indeed only have an effect if single and non-resonant contributions become
sizable.
\section{Summary}
We presented some selected results from our recent comparison of theoretical
predictions for $pp\to t\bar{t}W$ in the multi-lepton decay channel. We find
that fixed-order full off-shell and on-shell $t\bar{t}W$ NLOPS predictions are
overall in good agreement with each other within the estimated theoretical
uncertainties. Nonetheless, parton-shower based predictions have considerable
shape differences in comparison to fixed-order approaches. 
The observed differences are enhanced in the case of the $t\bar{t}W$ EW
contributions, which however is itself of the order of $10\%$ of the leading
QCD prediction. Therefore, differences in the $t\bar{t}W$ EW predictions only
have a minor impact on the final predictions.

In the absence of NLOPS predictions for the full off-shell calculation we
proposed a simple combination procedure in order to improve on currently
available theoretical predictions. However, in the future NLOPS predictions for
the full off-shell calculation as well as predictions including NNLO QCD
corrections for on-shell $pp\to t\bar{t}W$ will become necessary.
\begin{description}
 \item[Acknowledgements:] This work is supported in part by the U.S. Department
 of Energy under the grand DE-SC010102.
\end{description}

\end{document}